\documentclass[aps,twocolumn,psfig]{revtex4} 

\usepackage{graphicx}
\usepackage{amsmath}
\usepackage{amsfonts}
\usepackage{amssymb}
\def\y{\'{\i}}

\def\bea{\begin{eqnarray}}
\def\eea{\end{eqnarray}}
\def\beq{\begin{equation}}
\def\eeq{\end{equation}}

\begin{document}
\vspace*{-1.cm}

\title{Importance of Granular Structure in the Initial Conditions for the Elliptic Flow} 
\author{R.P.G.~Andrade$^1$, F.~Grassi$^1$, Y. Hama$^1$, 
 T. Kodama$^2$, and W.L.~Qian$^1$} 

\vspace*{1.cm}

\affiliation{$^1$ Instituto de F\y sica, Universidade 
             de S\~ao Paulo, C.P. 66318, 05315-970 
             S\~ao Paulo-SP, Brazil \\ 
             $^2$ Instituto de F\y sica, Universidade 
             Federal do Rio de Janeiro, C.P. 68528, 
             21945-970 Rio de Janeiro-RJ , Brazil} 

\begin{abstract}
We show effects of granular structure of the initial 
conditions (IC) of hydrodynamic description of high-energy 
nucleus-nucleus collisions on some observables, especially 
on the elliptic-flow parameter $v_2$. Such a structure 
enhances production of isotropically distributed high-$p_T$ particles, making $v_2$ smaller there. Also, it reduces $v_2$ in the forward and backward regions where the global matter density is smaller, so where such effects become more  efficacious. 
\vspace*{.3cm}

\noindent PACS numbers: 25.75.-q,24.10.Nz,25.75.Ld  
\end{abstract}

\maketitle

\section{Introduction} 

It is by now widely accepted that hydrodynamics is a 
successful approach for describing the collective 
flow in high-energy nuclear collisions. 
The basic assumption in hydrodynamical models is the 
local thermal equilibrium. It is assumed that, after a complex process involving microscopic collisions of nuclear constituents, at a certain early instant a hot 
and dense matter is formed, which would be in local 
thermal equilibrium. 
Usually, this state is characterized by some initial  conditions (IC), parametrized as smooth distributions of thermodynamic quantities and four-velocity. 
After this instant, the system 
would evolve hydrodynamically, following the well 
known set of differential equations. 

However, since our systems are not large, important  event-by-event fluctuations are expected in real collisions.  Concerning this question, fluctuation in the IC deserves a special consideration. 
In the past few years, we have studied several effects  caused by such fluctuating IC on observables, by using 
a computational code especially developed for this purpose, which we call NeXSPheRIO \cite{preliminary,hbt,b_part,qm05,v2}. In particular, in \cite{preliminary} we have studied the fluctuations of the so called elliptic flow parameter $v_2$, showing that they are quite large, which has been confirmed by recent data \cite{STAR,PHOBOS}. Our more recent  computations gave similar results \cite{fv2}. 

An important point with regard to such IC for high-energy nuclear collisions is that they are not only event-by-event fluctuating, but also are strongly inhomogeneous in space. 
Since the incident nuclei are not smooth objects, if the thermalization is verified at very early time as usually assumed in hydrodynamic approach, {\it they could not be smooth but should have granular structure}. 
In the present work, we focus our attention mainly to such a 
{\it granular structure} of fluctuating IC and try to show 
important effects on some observables, especially on the elliptic-flow parameter $v_2$. 

In what follows, we will first give a brief discussion on 
{\it what is expected if such hot blobs are produced}. This 
will be done in the next Section. 
We shall then describe in Section~\ref{NS}, our main tool,  the NeXSPheRIO code. 
We show the results of computations in Section~\ref{results}, 
first on transverse-momentum ($p_T$) spectra, and then on $v_2\,$ as function both of pseudo-rapidity $\eta$ and of  
$p_T\,$. 
Finally, main conclusions are drawn in Sec.\ref{conclusions}.

\section{What is expected from the hot blobs?} 
 \label{hot blobs} 

\begin{figure}[htb]
\vspace*{-.3cm} 
\begin{center}
\includegraphics*[width=8.cm]{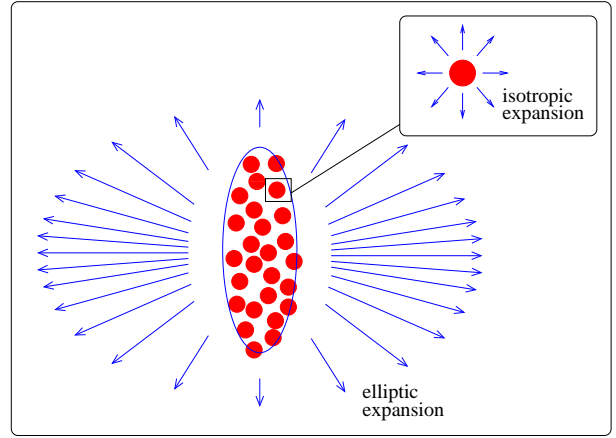}
\end{center} 
\caption{Pictorial representation of the energy density  
 distribution in a fluctuating IC.} 
\label{blobs}
\end{figure} 

What do we expect if the IC present granular structure as depicted in Fig.$\,$\ref{blobs}$\,$? Because of high  concentration of energy in {\it point-like} regions, we  imagine that initially each blob would suffer a violent explosion and, because of their small size, isotropically. 
If one of such blobs is deep inside the hot matter, this  initial motion is quickly absorbed by the surrounding medium, so would not result in any observable effect. However, if  such a blob is at the surface of the matter, certainly the  outgoing part of this initial acceleration would remain producing high-$p_T$ particles, which would be isotropically distributed in the momentum space. 

Thus, first we expect that high-$p_T$ part of the $p_T$ spectra is enhanced when fluctuating IC are used in our computations, in comparison with the results with averaged (smooth) IC. In the second place, we expect that the 
anisotropic-flow coefficient $v_2$ suffers reduction as we 
go to high-$p_T$ region, due to the additional high-$p_T$ 
isotropic component  included now. 
As for the $\eta$ dependence of $v_2\,$, we know that the average matter density decreases as $\vert\eta\vert$ increases as reflected in the $\eta$ distribution of charged particles, so when such a blob is formed in the  large-$\vert\eta\vert$ regions, its effects appear more clearly. Therefore, we expect considerable reduction of $v_2\,$ in those regions. 

Although not discussed here, the granular structure we are 
considering certainly affect the so-called HBT radii. This question has been discussed in a previous publication \cite{hbt}. 
\vspace*{-.2cm} 

\section{NeXSPheRIO Code}
 \label{NS} 
 
Our fundamental tool for the present study is called NeXSPheRIO. It is a junction of two computational codes:  NeXus and  SPheRIO. The NeXus code \cite{nexus} is used to compute the IC:  $T^{\mu \nu}$ and $j^{\mu}$ on some initial  hypersurface. It is a microscopic model based on the Regge-Gribov theory and the main advantage for our purpose is that, once a pair of incident nuclei or hadrons and their incident energy are chosen, it can produce, in the  event-by-event basis, detailed space distributions of  energy-momentum tensor, baryon-number, strangeness and charge densities, at a given initial time  $\tau=\sqrt{t^2-z^2}\sim1\,$fm. 
Remark that, when we use a microscopic model to create 
a set of IC for hydrodynamics, the generated energy-momentum 
tensor does not necessarily correspond to that of local 
equilibrium, so we need to transform it to that of the 
equilibrated  matter, adopting some procedure as described in 
detail in Ref. \cite{review}. 

\begin{figure}[h] 
\begin{center}
\includegraphics*[angle=-90, width=8.2cm]{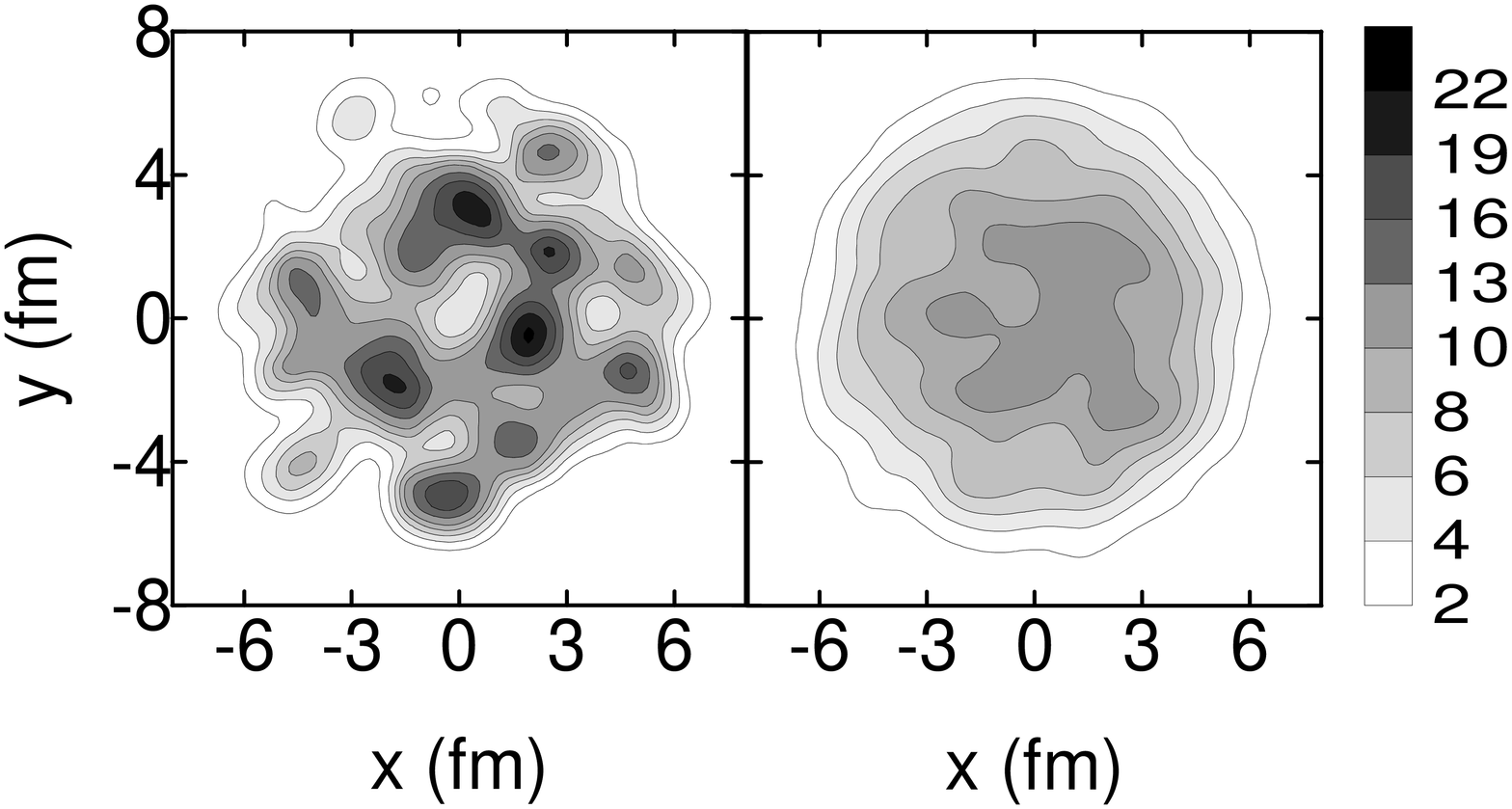}
\end{center} 
\vspace*{-.3cm} 
\caption{Examples of initial conditions for 
central Au+Au collisions given by NeXus at mid-rapidity 
plane. The energy density is plotted in units of 
GeV/fm$^3$. 
Left: one random event. Right: average over 30 random 
events (corresponding to the smooth initial conditions 
in the usual hydro approach).} 
\label{ic}
\end{figure}

We show in Fig.~\ref{ic} an example of such a 
fluctuating event, produced by NeXus event generator, 
for central Au~+~Au collision at $130\,A\,$GeV, compared with an average over 30 events. As can be seen, the 
energy-density distribution for a single event (left), 
at the mid-rapidity plane, presents several blobs of 
high-density matter, whereas in the averaged IC (right) the distribution is smoothed out, even though the number of events is only 30. The latter would corresponds to the usually adopted smooth and symmetrical IC in many hydrodynamic calculations. 
The bumpy event structure, as exhibited in Fig.~\ref{ic}, was also shown in calculations with {HIJING}~\cite{gyulassy} and other event generators. As already observed there and studied in \cite{preliminary,hbt,b_part,qm05,v2,review}, this bumpy structure gives important consequences in the observables. 
As for the velocity distribution, it gives essentially zero 
transverse velocity and longitudinal component close to the 
boost-invariant one. 

Solving the hydrodynamic equations for events, so 
irregular as the one shown in Fig~\ref{ic}, 
requires a special care. The SPheRIO code is well 
suited to computing the hydrodynamic evolution of 
such systems. It is based on Smoothed Particle 
Hydrodynamics (SPH), an algorithm originally developed in 
astrophysics \cite{sph} and adapted to relativistic 
heavy ion collisions \cite{sp1}. 
It parametrizes the flow in terms of discrete  Lagrangian coordinates attached to small volumes  (called ``particles'') with  some conserved quantities. 
Its main advantage is that any geometry in the initial 
conditions can be incorporated and this is done giving a desired precision. 

Now, we have to specify some equation of state (EoS) 
describing the locally equilibrated matter. Here, 
in accordance with Ref.~\cite{qm05}, we will adopt 
a phenomenological implementation of EoS, giving 
a critical end point in the quark-gluon plasma - hadron gas transition line, as suggested by the lattice QCD~\cite{LQCD}. 

We shall neglect in this paper any dissipative effects and also assume the usual sudden freezeout at a constant  temperature. As for the conserved quantities, besides the  energy, momentum and entropy, we consider just the baryon  number. Although very simplified, we believe that the main  
outcomes of the present study will remain valid in more  detailed description. 

In computing several observables, we perform in the present work two sets of computations: i) First, we average over  random NeXus events, obtaining smooth IC, which are used to compute the observables by using the SPheRIO code. This is  similar to the usual hydro calculation. 
ii) In the second set, NeXSPheRIO is run many times and an 
average over final results is performed. This mimics experimental conditions more closely. Remark that in the latter the granular structure of IC, mentioned above, are being explicitly included whereas in the former not. 

Having depicted our tool, let us now explain how we fix the parameters of the model and compute the observables of our interest. 
Certainly any model to be considered as such should reproduce the most fundamental, global  quantities involving the class of phenomena for which it is proposed. So, we begin by fixing the initial conditions so as to reproduce properly the 
(pseudo-)rapidity distributions of charged particles 
in each centrality window. 
This is done by applying an $\eta$-dependent factor 
$\sim1$ to the initial energy density distribution of all the events of each centrality class, produced by NeXus. 
Next, we would like to correctly reproduce the  transverse-momentum spectra of charged particles, 
which can be achieved by choosing an appropriate centrality 
dependent freezeout temperature, $T_{fo}\,$. 
\vspace*{-.2cm} 

\section{Results} 
 \label{results} 
 
\subsection{Charged-particle spectra} 

\begin{figure}[h!tb]
\begin{center}
\includegraphics*[width=8.cm]{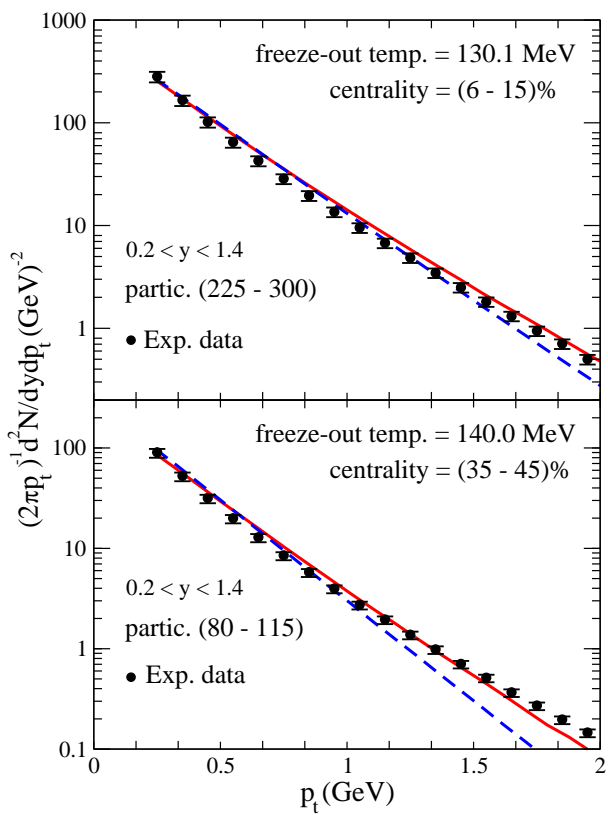}
\end{center} 
\vspace*{-.4cm} 
\caption{Charged-particle $p_T$ distributions, in two 
 different centrality windows, computed in two different ways 
 as explained in the text. The solid lines indicate results 
 for event-by-event fluctuating IC, whereas the dotted lines 
 the ones for the averaged IC. The data points are also 
 plotted for comparison \cite{PHOBOS3}.} 
\label{dndpt}
\end{figure}

We show in Fig.$\,$\ref{dndpt}$\,$, our results for the charged-particle spectra in two different centrality windows as indicated, computed as explained in Section \ref{NS}. 
It is clearly seen that, as one goes from the smooth averaged IC to bumpy fluctuating IC, the high-$p_T$ component of the spectra increases as expected, making them more concave and closer to the data. 

\vspace*{-.5cm} 

\subsection{$p_T$ dependence of $\langle v_2\rangle$} 

\begin{figure}[h]
\begin{center}
\vspace*{-.2cm} 
\includegraphics*[width=8.cm]{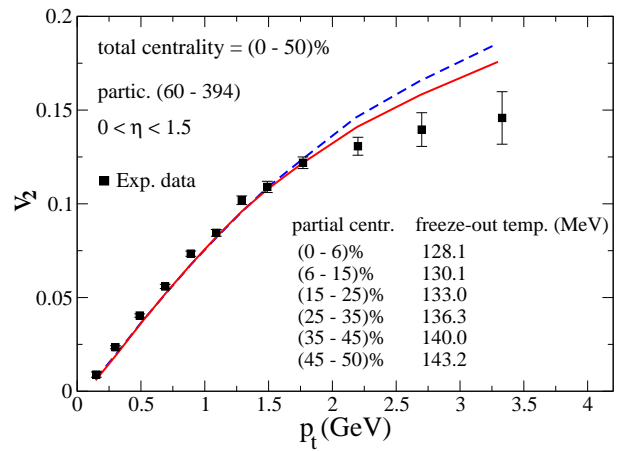}
\end{center} 
\vspace*{-.5cm} 
\caption{$p_T$ dependence of $\langle v_2\rangle$ in the 
 centrality window and $\eta$ interval as indicated, compared 
 with data \cite{PHOBOS4}. The 
 solid line indicates result for fluctuating IC, whereas 
 the dotted one that for the averaged IC. The curves are  
 averages over PHOBOS centrality sub-intervals with  
 freeze-out temperatures as indicated. 
 } 
\label{v2pt}
\end{figure}

In Fig.$\,$\ref{v2pt}$\,$, we show our results for the $p_T$ dependence of $\langle v_2\rangle$ in the centrality window 
and $\eta$ interval as indicated. Here, since our purpose is 
to show clearly the effects of granular IC, we plot the 
results obtained with the same PHOBOS hit-based method for 
the both curves, but without any correction for event-plane fluctuations. While the result with smooth averaged IC shows 
continuous increase of $\langle v_2\rangle$ with  $p_T\,$, 
deviating largely from the data points, as happens in the 
orthodox hydro  computations, the introduction of spiky IC 
makes $v_2$ smaller at high $p_T$ as expected, approaching 
the curve to the data points. 
The correction mentioned above shifts upward the curve, 
corresponding to the fluctuating IC, but without modifying its shape. 
The fact that the averaged smooth IC lead to rising $v_2(p_T)$ for large $p_T$ and not flattening is usually interpreted as a breakdown of hydrodynamic model. We suggest it could be in part related to the granular structure of the IC. 

\vspace*{-.5cm} 

\subsection{$\eta$ dependence of $\langle v_2\rangle$} 

\begin{figure}[h]
\begin{center}
\includegraphics*[width=8.cm]{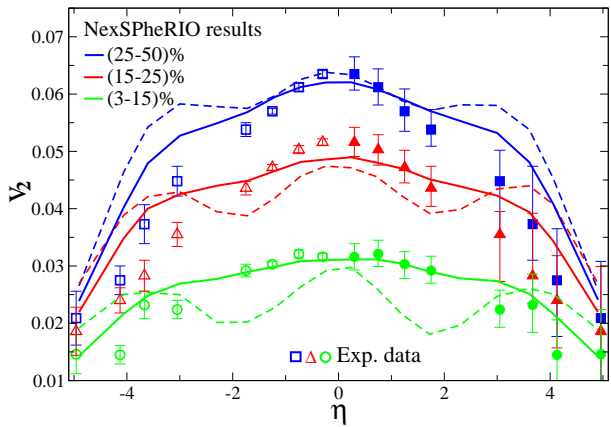}
\end{center} 
\vspace*{-.5cm} 
\caption{$\eta$ dependence of $\langle v_2\rangle$ for three 
 centrality windows. The solid lines indicate results 
 for event-by-event fluctuating IC, whereas the dotted lines 
 the ones for the averaged IC. The data points are also 
 plotted for comparison \cite{PHOBOS4}. $T_{fo}$ has been 
 taken as indicated in FIG.\ref{v2pt}.} 
\label{v2eta}
\vspace*{-.4cm}
\end{figure}

In Fig.$\,$\ref{v2eta}$\,$, we show our results for the $\eta$ dependence of $\langle v_2\rangle$ in three different  centrality windows as indicated. 
As in Ref. \cite{v2}, we calculated $\langle v_2\rangle$ with respect to the event plane as done experimentally. The event plane has been determined here by all the charged particles  (the results with the PHOBOS hit-based method with correction are almost identical). For averaged smooth IC, in agreement with 
the usual hydro computations \cite{hirano1,hirano2,nonaka}, 
$\langle v_2\rangle$ exhibits shoulders in 
high-$\vert\eta\vert$ regions. 
When fluctuating spiky IC are used, these shoulders are 
considerably weakened as expected. 
The results, which combine this and the complementary effect of increase of $\langle v_2\rangle$ with the event-plane fluctuations, are now very close to the data.

\medskip 

Another important ingredient of a hydrodynamic model is the 
decoupling procedure. We are studying the effect of 
{\it continuous emission} \cite{ce}. Probably it makes the curve of $v_2$ at high-$p_T$ even flatter and the $\eta$-distribution narrower \cite{qm05}. 
This is clear, also in Refs. \cite{hirano2,nonaka,hirano3},  where they use the usual smooth IC and, by describing the  decoupling with a microscopic transport model, could 
improve $v_2$ in this direction. 
The cascade model in the hadronic stage produces similar effects as the 
continuous emission. 

\section{Conclusions} 
 \label{conclusions} 

While it is believed that a granular structure exists in the IC, hydro simulations are usually done with smooth IC, the expectation being that the granularity will not manifest itself. In this paper, we argue that this may not be true qualitatively and also quantitatively. 
\smallskip 

The main conclusions of the present study are: 

\begin{enumerate} 
 \item 
 Granular structure in the IC produces more high-$p_T$  
 particles, distributed isotropically.  
 \item
 Granular structure in the IC reduces the elliptic flow, 
 because of the isotropic flow it generates. 
 \item 
 As function of $p_T$, the mechanism becomes more effective 
 as $p_T$ increases, because those high-density blobs cause 
 violent expansion, producing high-$p_T$ particles. 
 \item 
 This effect is enhanced where the average matter density is  
 small. So, it decreases $v_2$ in the large pseudorapidity  
 regions. 
 \item 
 NeXSPheRIO, with fluctuating and spiky IC, 
 reproduce approximately the 
 data both of the $p_T$-distribution and 
 $\eta$-distribution of $v_2$ for different centrality 
 windows. 
\end{enumerate}

\noindent{\bf Acknowledgments:} The authors thank Roy Lacey for the initial discussions, which motivated this study.  This work has been financially supported by FAPESP, CNPq,  FAPERJ and PRONEX.

\end{document}